\documentclass[aps,pre,twocolumn,showkeys,superscriptaddress]{revtex4-1}

\usepackage{amsfonts}
\usepackage{amsmath}
\usepackage{amssymb}
\usepackage{graphicx}
\usepackage{color}
\usepackage{amsmath}
\usepackage{mathtools}
\usepackage{hyperref}
\usepackage{float}
\usepackage{cancel}
\usepackage{upgreek}
\usepackage{mathbbol}

\newcommand{\thickness}{w}


\begin{document}

\title{The elastic inclusion problem in the (amplitude) phase field crystal model}

\author{Marco Salvalaglio}
\affiliation{Institute of Scientific Computing, TU Dresden, 01062 Dresden, Germany}
\affiliation{Dresden Center for Computational Materials Science, TU Dresden, 01062 Dresden, Germany}

\author{Karthikeyan Chockalingam}
\affiliation{Science and Technology Facilities Council, Daresbury Laboratory, Daresbury Science and Innovation Campus, Warrington, Cheshire WA4 4AD, UK}

\author{Axel Voigt}
\affiliation{Institute of Scientific Computing, TU Dresden, 01062 Dresden, Germany}
\affiliation{Dresden Center for Computational Materials Science, TU Dresden, 01062 Dresden, Germany}

\author{Willy D\"orfler}
\affiliation{Institute for Applied and Numerical Mathematics, Karlsruhe Institute of Technology, 76131 Karlsruhe, Germany}


\begin{abstract}
In many processes for crystalline materials such as precipitation, heteroepitaxy, alloying, and phase transformation, lattice expansion or compression of embedded domains occurs. This can significantly alter the mechanical response of the material. Typically, these phenomena are studied macroscopically, thus neglecting the underlying microscopic structure. Here we present the prototypical case of an elastic inclusion described by a mesoscale model, namely a coarse-grained phase-field crystal model. A spatially-dependent parameter is introduced into the free energy functional to control the local spacing of the lattice structure, effectively prescribing an eigenstrain. The stress field obtained for an elastic inclusion in a 2D triangular lattice is shown to match well with the analytic solution of the Eshelby inclusion problem.
\end{abstract}


\maketitle

\section{Introduction}

The study of inclusions in crystalline materials is of great importance for many materials science and engineering applications. For instance, this applies to processes such as precipitation, phase-transformation, heteroepitaxy \cite{li2008introduction,cai2016imperfections,Bergamaschini2016b}, often with technological relevance and also involving additional aspects such as capillarity \cite{muller2004elastic}. A prominent example in this context consists of phase changes introducing structural transformation of the host lattice, which might even affect the lattice symmetry. This occurs, for instance, in lithium-ion batteries, where the flow of lithium into the electrode particle introduces lattice expansion of the host material. The size of these systems is typically in the order of 1 $\mu m$. Therefore, continuum methods generally are adopted. For instance, the numerical modeling of phase transformation in lithium-ion batteries has been tackled by a classical phase-field approach coupled to elasticity to account for the mechanical equilibrium of elastic deformation~\cite{Huttin2012}.

Continuum approaches are powerful in describing crystalline systems at the macroscale. However, they neglect details of the microscopic scales such as crystal symmetries, anisotropies, and orientation of the grains in polycrystalline materials. Methods capable of tackling the resulting effects of these microscopic details in a macroscopic description are highly demanded to obtain comprehensive descriptions. In this paper, we present the modeling of an elastic inclusion by the phase-field crystal (PFC) model \cite{Elder2002,Elder2004,Emmerich2012}, focusing in particular on its amplitude expansion (APFC) formulation \cite{Goldenfeld2005,Athreya2006,YHET:2010,SE:2022}. This approach allows for describing elasticity on a microscopic scale \cite{ElderPRE2010,SpatschekKarma:2010,Heinonen2014,Huter2016} while bridging the gap among micro- and macro-scale descriptions of crystal structures under some approximations \cite{SalvalaglioNPJ2019,SAHVEV:2020,SalvalaglioPRL2021}. 
We formulate the problem incorporating a prescribed lattice expansion/compression in the free energy through a spatial dependent parameter that controls the local lattice spacing. We show that the model reproduces the stress field of a spherical inclusion, thus encoding an eigenstrain formulation \cite{Kinoshita1971}. The approach retains details of the underlying lattice structure as conveyed by the APFC model. An example for 2D crystals with triangular symmetry is explicitly given, which can be compared to analytic solutions for the elastic inclusion, i.e., with the Eshelby problem \cite{eshelby1957determination,eshelby1959elastic,mura2013micromechanics}. This comparison serves as a proof of concept for more general elastic inclusion problems.

\section{Amplitude phase-field crystal modeling}
\label{sec:apfc}

The PFC model describes the crystal lattices by means of a continuous, periodic order parameter $n:\Omega \rightarrow \mathbb{R}$, $\mathbf{r}\mapsto n(\mathbf{r})$, representing an atomic probability density \cite{Elder2002,Elder2004,Emmerich2012}.
The model is based on the free energy 
\begin{equation}
\begin{split}
   \label{eq:F_1}
    F(n) \!=\! \int_\Omega
      \bigg[ & \frac{B_0^x}{2}\big((q_0^2+\nabla^2)n\big)^2 \\
   & + \frac{\Delta B_0}{2}n^2 - \frac{\tau}{3}n^3+\frac{v}{4}n^4\bigg]{\rm d}\mathbf{r}
    \end{split}
\end{equation}
and an associated conserved gradient flow
\begin{equation}
\frac{\partial n}{\partial t} = \nabla^2 \frac{\delta F}{\delta n}, 
\label{eq:evo_n}
\end{equation}
with $t$ the time parameter. The parameter $q_0$ sets the periodicity of $n(\mathbf{r})$ and it is generally inversely proportional to the lattice spacing $a_0$. $B_0^x$, $\Delta B_0$, $\tau$, $v$ are real parameters as in Refs.~\cite{elder2007,ElderPRE2010} controlling which phase described by $n$ minimizes the energy and elastic properties. The order parameter $n$ can be well described by a sum of plane waves
\begin{equation}
n(\mathbf{r})=n_0+\sum_j^N \eta_j e^{\mathbb{i}\mathbf{q}_j \cdot {\mathbf{r}}} + {\rm c.c.},
\label{eq:ansatz_apfc}
\end{equation}
with c.c. denoting the complex conjugate, accounting for the contribution of $-\mathbf{q}_j$ for which $\eta_{-\mathbf{q}_j}=\eta_{\mathbf{q}_j}^*$ being $n(\mathbf{r})\in \mathbb{R}$, $n_0$ the average density, $\eta_j$ amplitudes, $\mathbb{i}$ the imaginary unit, and $\{\mathbf{q}_j\}$ a set of reciprocal lattice vectors encoding the symmetry of the crystal. 

An eigenstrain \cite{Kinoshita1971} encoding a lattice deformation from a lattice parameter $a_{0}$ to a lattice parameter $a_{\rm d}$ may be defined as $\varepsilon^* = (a_{\rm d}-a_{0})/a_{0} = q_0 / q_{\rm d} -1$.
Therefore, one may express the change encoded by $\varepsilon^*$ as 
\begin{equation}
    q_d = \frac{q_0}{1+\varepsilon^*} = \beta q_0,
    \label{eq:beta}
\end{equation}
with $\beta =1/(1+\varepsilon^*)$. Notice that $\beta =1$ corresponds to $\varepsilon^* =0$, while $\beta <1$ and $\beta >1$ correspond to positive or negative eigenstrains, respectively. Moreover, $\beta$ diverges only in the unphysical limit $a_{\rm d} \rightarrow 0$. No restrictions exist to consider this parameter spatially dependent, namely $\beta(\mathbf{r}) =1/(1+\varepsilon^*(\mathbf{r}))$. To encode an eigenstrain $\varepsilon^*$ in the PFC model we consider then a modified energy functional featuring a slowly varying quantity $\beta(\mathbf{r})$,
\begin{equation}
\begin{split}
   \label{eq:F_2}
    F_\beta(n) = \int_\Omega 
      \bigg[& \frac{B_0^x}{2}\left((\beta(\mathbf{r})^2q_0^2+\nabla^2)n\right)^2 \\ &
    + \frac{\Delta B_0}{2}n^2 - \frac{\tau}{3}n^3+\frac{v}{4}n^4\bigg]{\rm d}\mathbf{r}.
    \end{split}
\end{equation}

The PFC model naturally accounts for elasticity \cite{Elder2002,Elder2004}. Elastic effects can be characterized by focusing on a small perturbation of the density in eq.\ \eqref{eq:ansatz_apfc} due to a displacement field $\mathbf{u}$. As a result, elasticity effects may be fully described by complex amplitudes $\eta_j=\phi_j e^{\mathbb{i}\mathbf{q}_j\cdot \mathbf{u}}$ with $\phi_j$ the (real) amplitudes for a relaxed crystal \cite{ElderPRE2010,Heinonen2014,Huter2016}. In the amplitudes expansion of the PFC model, the APFC model, $\boldsymbol{\eta} =[\eta_j]_{j=1,\dots,N}$ are the variable to solve for \cite{Goldenfeld2005,Athreya2006,YHET:2010,SE:2022}. They are associated to a minimal set of $N$ reciprocal lattice vectors that describes a targeted lattice symmetry entering eq.\ \eqref{eq:ansatz_apfc}.
This approach allows for coarse-grained description of the lattice that approaches macroscopic lengthscales but still retaining microscopic details \cite{SalvalaglioNPJ2019,SAHVEV:2020}. 
The corresponding equation may be derived by 
substituting the ansatz in eq.\ \eqref{eq:ansatz_apfc} in eq.\ \eqref{eq:F_2} and integrating over the unit cell. This procedure may be rigorously justified by multiple scales expansions or renormalization group calculations \cite{Goldenfeld2005,Athreya2006}. The APFC free energy obtained by this procedure, with $n_0=0$ without loss of generality, reads 
\begin{equation}
\begin{split}
   \tilde{F}_\beta(\boldsymbol{\eta})
    = \int_\Omega  \Bigg[ & 
\sum_{j=1}^{N} \bigg( B_0^x 
      \left|\mathcal{G}_j\eta_j\right|^2 - \frac{3v}{2} \left|
      \eta_j\right|^4 \bigg)+\\   &
    \frac{\Delta B}{2} \Phi
      + \frac{3v}{4}\Phi^2 + f^{\textrm{s}}(\boldsymbol{\eta},\boldsymbol{\eta}^*)  
      \Bigg]{\rm d}\mathbf{r},
      \end{split}
      \label{eq:F_eta}
\end{equation}
where $\Phi\equiv 2\sum_{j=1}^{N}\left|\eta_j\right|^2$ and 
\begin{equation}
   \label{eq:def_cG}
   \mathcal{G}_j(\mathbf{r})
      \equiv \nabla^2 + 2\mathbb{i}\mathbf{q}_j\cdot\nabla
      + \beta(\mathbf{r})^2-1.
\end{equation}
$f^{\textrm{s}}$ is a polynomial which takes different forms according to the considered symmetry \cite{SE:2022,ElderPRE2010,SBVE:2017}. Here we consider 2D crystals with triangular symmetry described in a one-mode approximation (considering the shortest reciprocal lattice vectors only), i.e. $N=3$:
$\mathbf{q}_1=  q_0\left(-\sqrt{3}/2,-1/2 \right)$, $\mathbf{q}_2= q_0(0,1)$, 
$\mathbf{q}_3= q_0\left(\sqrt{3}/2,-1/2 \right)$ with $q_0=1$ and
\begin{equation}
f^{\rm tri}=-2\tau(\eta_1\eta_2\eta_3+\eta_1^*\eta_2^*\eta_3^*).
\label{eq:energyterm_tri}
\end{equation}
With these choices, $a_0=4\pi/\sqrt{3}$. The dynamics of ${\eta_j}$ obtained from the dynamics of $n$, eq.\ \eqref{eq:evo_n} with a procedure similar to the derivation of the energy $\tilde{F}_\beta$ \cite{SE:2022} reads
\begin{equation}
\begin{split}
   \label{eq:evolution-eta}
   \frac{\partial \eta_j}{\partial t} = &-|\mathbf{q}_j| \frac{\delta\tilde{F}_\beta}{\delta\eta_j^*}\\
   =& -\Big(\Delta B_0 + B_0^x\mathcal{G}_j^2 + 3v(\Phi - \left|\eta_j \right|^2) \Big)\eta_j
     - \frac{\partial f^{\rm s}}{\partial \eta_j^*}.
   \end{split}
\end{equation}



Minimizers of $\tilde{F}_\beta$ denote equilibrium configurations.
A relaxed crystal, corresponding to the lattice represented by $\{\mathbf{q}_j\}$ is described by real, constant amplitudes, which take some values depending on the length of the corresponding $\mathbf{q}_j$ vectors \cite{ElderPRE2010,SBVE:2017}. If we assume $\eta_j =\phi$ for some real $\phi$ we obtain
\begin{equation*}
   \tilde{F}_\beta(\phi)
   = \int_{\Omega} \bigg[3(\Delta B_0+(\beta^2-1)^2)\phi^2 + \frac{45v}{2}v\phi^4  - 4\tau\phi^3 \bigg] {\rm d} \mathbf{r}.
\end{equation*}
This energy is minimized by
\begin{align}
   \label{eq:def_phi_+}
   \phi_\pm = \frac{\tau \pm \sqrt{\tau^2-15v (\Delta B_0+(\beta^2-1)^2)}}{15v}.
\end{align}
Here we will look at $\phi_{+}$ by restricting our analysis to $t>0$ without loss of generality. Real solutions thus exist if $\Delta B_0 <(\tau^2/15v)-(\beta^2-1)^2$, while the solid phase is favored if $\Delta B_0 <8\tau^2/135v-(\beta^2-1)^2$ and $\Delta B_0 =8\tau^2/135v-(\beta^2-1)^2$ is the solid/liquid or ordered/disordered coexistence condition. For $\beta =1$ we recover the conditions given in \cite{SBVE:2017}.


From the energy in eq.\ \eqref{eq:F_2} one can also derive the stress field $\boldsymbol{\sigma}^{n}$ \cite{Skaugen2018,Skaugen2018b,SAHVEV:2020,SSAV:2021}. In our case, considering a slowly varying inhomogeneous $\beta$, we obtain
\begin{equation}
   \boldsymbol{\sigma}^n
   = 2{\rm Sym}\bigg( 
      \nabla\big( [\nabla^2 + \beta(\mathbf{r})^2]n \big)
      \otimes\nabla n \bigg),
      \label{eq:sigma-psi-beta}
\end{equation}
where we have omitted the isotropic pressure term due to negligible contribution. Inserting now the amplitude ansatz from eq.\ \eqref{eq:ansatz_apfc}, leads to an amplitude depending deformation gradient $\boldsymbol{\sigma}^{\boldsymbol{\eta}}$, given by 
\begin{equation}
\begin{split}
   \sigma_{lm}^{\boldsymbol{\eta}}
   = \sum_{j=1}^N &
     \bigg( (\partial_l+\mathbb{i} q_l^j)\mathcal{G}_{j}\eta_{j}\,
     (\partial_m-\mathbb{i} q_m^j)\eta_{j}^*+\\ &(\partial_m+\mathbb{i} q_m^j)\mathcal{G}_{j}\eta_{j}\,
    (\partial_l-\mathbb{i} q_l^j)\eta_{j}^*\bigg),
    \end{split}
    \label{eq:sigma-eta-beta}
\end{equation}
for $l,m\in\{1,2\}$, where $\mathcal{G}_j$ is defined in eq.\ \eqref{eq:def_cG}, recovering the expressions in \cite{SAHVEV:2020} for $\beta=1$. Eq.\ \eqref{eq:sigma-eta-beta} is expected to deliver stress fields accounting for non-linearities and strain-gradient terms \cite{Huter2016,SAHVEV:2020}.

\section{The Elastic inclusion problem}
\label{sec:eshelby}

The calculation of the stress/strain field in the presence of an elastic inclusion, namely a portion of a material with an eigenstrain $\varepsilon^*$ surrounded by a relaxed medium, is known as Eshelby's inclusion problem \cite{eshelby1957determination,eshelby1959elastic,mura2013micromechanics}.
The original formulations focused on the elastic field in the inclusion and involved the assumption of an infinite matrix surrounding it. Following works focused on the derivation of the solution addressing finite systems with specific boundary conditions \cite{mura2013micromechanics,ju1999novel,li2005circular,wang2005circular}. For the example delivered in this work, we consider explicitly the analytic solutions obtained in an infinite medium and will comment on the comparisons with simulations in the following. 

The stress tensor in the presence of an inclusion with eigenstrain matrix $\varepsilon^*_{kl}$ can be expressed as
\begin{equation}
   \sigma_{ij}^\mathrm{e} = \mathcal{C}_{ijkl} \varepsilon^{\mathrm{e}}_{kl}
      = \mathcal{C}_{ijkl} \big( \mathcal{S}_{klpq} \varepsilon^*_{pq}
        - \chi(\mathbf{r})\varepsilon^*_{kl}\big),
        \label{eq:eshelby}
\end{equation}
with $\mathcal{C}_{ijkl}$ the rank-four elasticity tensor, $\mathcal{S}_{klpq}$ the Eshelby tensor,  $\varepsilon^{\mathrm{e}}_{pq}$ the elastic strain tensor and $\chi(\mathbf{r})$ an indicator/characteristic function which is 1 in the inclusion and 0 outside. The deformation leading to a change in the lattice parameter translates to a diagonal eigenstrain matrix $\varepsilon^*_{kl}=\varepsilon^* \delta_{kl}$.
The elasticity tensor for an isotropic medium is expressed as 
\begin{equation}
   \mathcal{C}_{ijmn}
   = \lambda \delta_{ij} \delta_{mn}
     + \mu \big(\delta_{im}\delta_{jn} + \delta_{in}\delta_{jm} \big),
\end{equation}
where $\lambda$ and $\mu$ are material parameters (the Lam\'e constants). For the APFC description considered in Sect.~\ref{sec:apfc}, using eq.\ \eqref{eq:def_phi_+}, they read $\lambda = \mu = 3\phi_+^2$ \cite{Heinonen2014,Skaugen2018}.

The analytical solutions for the Eshelby tensor $\mathcal{S}^{\infty}$ of an inclusion embedded in an infinite medium at the interior (I) and exterior (E) to the inclusion reads \cite{mura2013micromechanics,ju1999novel,li2005circular},  $\mathcal{S}_{ijmn}^{\infty}=\chi(\mathbf{r})\mathcal{S}^{{\rm I},\infty}_{ijmn}+(1-\chi(\mathbf{r}))\mathcal{S}^{{\rm E},\infty}_{ijmn}$ with
\begin{equation*}
 \mathcal{S}^{{\rm I},\infty}_{ijmn} = \frac{3-4\nu}{8(1-\nu)}(\delta_{im}\delta_{jn}+\delta_{ij}\delta_{mn}) +  \frac{(4\nu-1)}{8(1-\nu)}\delta_{im}\delta_{jn},
 \end{equation*}
 and
 \begin{equation*}
\begin{split}
 \mathcal{S}^{{\rm E},\infty}_{ijmn}=&\frac{\rho^2}{8(1-\nu)}\bigg[
(\rho^2+4\nu-2)\delta_{ij}\delta_{mn}+4(1-\rho^2)\delta_{ij}r_mr_n
 \\ &
+(\rho^2-4\nu+2)(\delta_{im}\delta_{jn}+\delta_{in}\delta_{jm})+4(1-2\nu-\\ &\rho^2)\delta_{mn}r_ir_j+
4(\nu-\rho^2)(\delta_{mn}e_je_n+\delta_{jm}e_ie_n+
\\&
\delta_{in}e_je_m+\delta_{jn}e_ie_m)+
8(3\rho^2-2)e_ie_je_me_n\Bigg],
\end{split}
\end{equation*}
with $\mathbf{r}=(x_1,x_2)$, $e_i(\mathbf{r}):=x_i/|\mathbf{r}|$, $\rho:=a/|\mathbf{r}|$,  $|\mathbf{r}|=\sqrt{x_1^2+x_2^2}$, $a$ is the radius of the inclusion and $\nu$ is the Poisson ratio (equal to 0.25 in the plane-strain settings \cite{Skaugen2018}). Equivalent formulations in terms of the stress field and elastic constants $\nu$ and $E=\mu(3\lambda+2\mu)/(\lambda+\mu)$ can be found in Ref.~\cite{fischer2018elastic}.

\section{Numerical APFC simulations}
\begin{figure}
\centering
\includegraphics[width=0.45\textwidth]{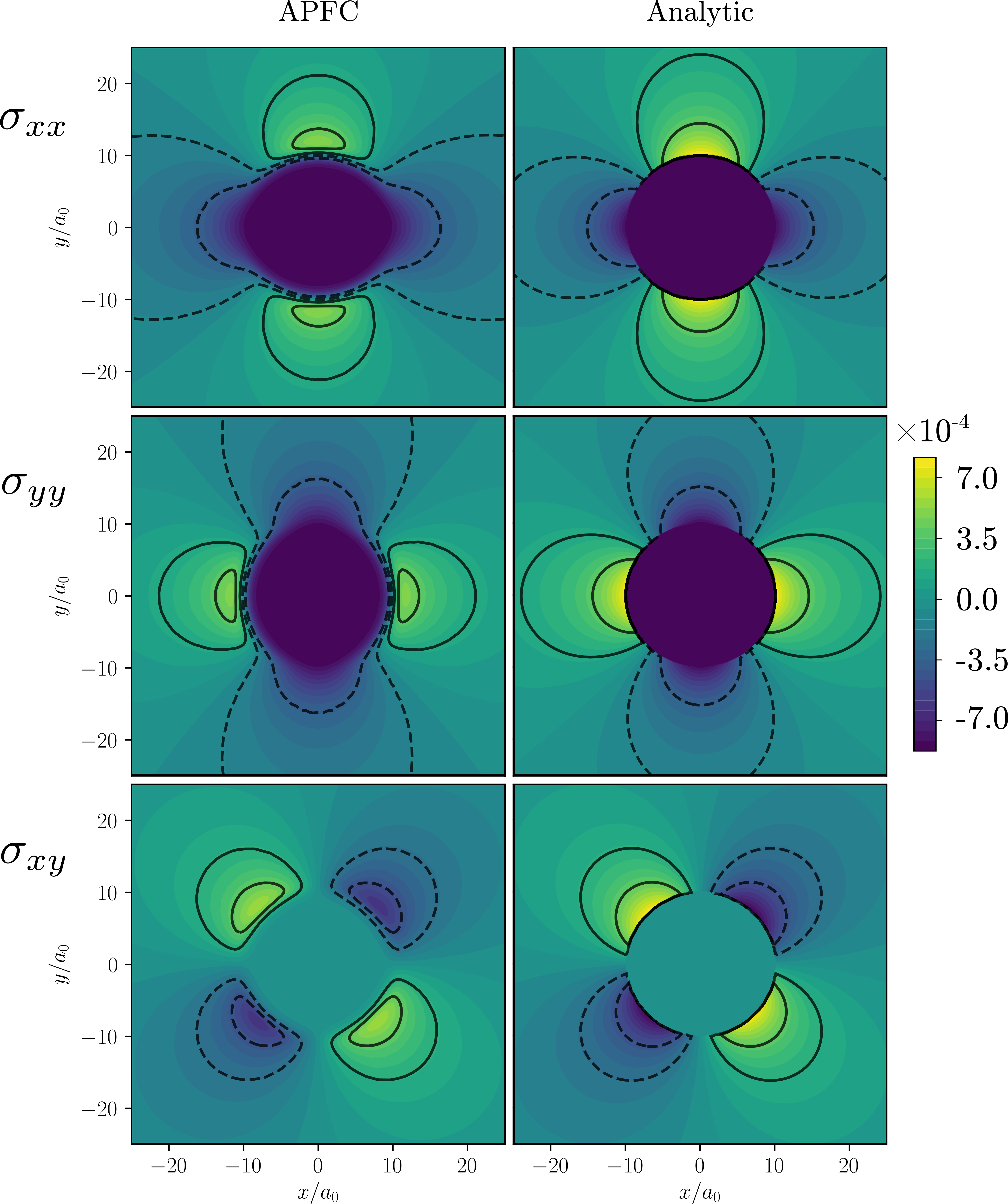}
  \caption{\label{fig:figure1}
Stress field components from APFC simulations (eq.\ \eqref{eq:sigma-eta-beta}, left column) and analytic solution for an inclusion in an infinite domain (eq.\ \eqref{eq:eshelby}, right column). Black contour lines are set to representative values for each stress component, and kept the same for both APFC and analytical stress field. $R=10a_0$, $\thickness=a_0$, $\varepsilon=0.01$ (other parameters are reported in the text).}
\end{figure}

In this section, we address the numerical simulation of the elastic inclusion problem within the APFC model illustrated in Sect.~\ref{sec:apfc} and discuss the results together with the analytic solution reported in Sect.~\ref{sec:eshelby}.

To deal with the continuous fields entering the (A)PFC models \footnote{With (A)PFC we refer to both PFC and APFC.}, the inclusion is described by a smooth approximation of the characteristic function $\chi$ considered in Sect.~\ref{sec:eshelby}. Defining a signed distance $d(\mathbf{r})$ from the boundary of the inclusion with negative sign in the inclusion and positive sign in the surrounding matrix, for a spherical inclusion with radius $R$ one obtains $d(\mathbf{r})=|\mathbf{r}|-R$. The characteristic function $\chi$ may then be approximated by $\chi_\thickness: \Omega \to [0,1]$ with
\begin{equation}
   \chi_\thickness(\mathbf{r}) := \frac{1}{2}
      \bigg[ 1-\text{tanh}\bigg( \frac{d(\mathbf{r})}\thickness \bigg)
   \bigg],
   \label{eq:tanh}
\end{equation}
which varies smoothly from $0$ in the matrix to $1$ inside the inclusion with $\thickness$ a parameter controlling the extension of the smoothing region. $\chi_\thickness(\mathbf{r})$ is used to set the lattice spacing in the inclusion and thus the eigenstrain therein. Using eq.\ \eqref{eq:beta} this is achieved by setting a spatial dependent $\beta(\mathbf{r})$ in eq.\ \eqref{eq:F_eta} as
\begin{equation}
   \beta(\mathbf{r})
   =  1 - \bigg(1 + \frac{q_d}{q_0}\bigg) \chi_\thickness(\mathbf{r}),
\end{equation}
which delivers $\beta q_0=q_0$ in the matrix, $\beta q_{\rm 0}=q_{\rm d}$ in the inclusion and provides an interpolation among these two values in between.

The APFC evolution equations \eqref{eq:evolution-eta} are solved exploiting the adaptive finite element toolbox AMDiS \cite{Vey2007,WitkowskiACM2015} with integration schemes as in Refs.~\cite{SBVE:2017,Praetorius2019} and minor adaptation to account for the $\beta(\mathbf{r})$ function. Further details concerning adaptive refinement, problem-tailored preconditioners and parallelisation strategies can be found therein. As initial condition we consider a spherical inclusion with radius $R=10a_0$ in a squared domain $100a_0 \times 100 a_0$. The model parameters are set to $\tau =1/2$, $v =1/3$, $B_0^x=1$, $\Delta B_0 =0.04$, the latter setting the system relatively close to the solid-liquid coexistence without loss of generality. Amplitudes are initialized to $\phi_+$ and the system in eq.\ \eqref{eq:evolution-eta} is allowed to relax until a steady state is reached. Periodic boundary conditions are used for all the amplitudes to consider the case usually adopted for APFC simulations. 

In Figure \ref{fig:figure1} the stress field obtained by evaluating eq.\ \eqref{eq:sigma-eta-beta} with $\boldsymbol{\eta}$ computed from APFC and $\thickness=a_0$, is compared with the analytic solution given in Sect.~\eqref{sec:eshelby}. The diffuse nature of the inclusion boundary encoded in eq.\ \eqref{eq:tanh} leads to a smooth field, still entailing the main features of the analytic solution.

\begin{figure}
\centering
\includegraphics[width=0.45\textwidth]{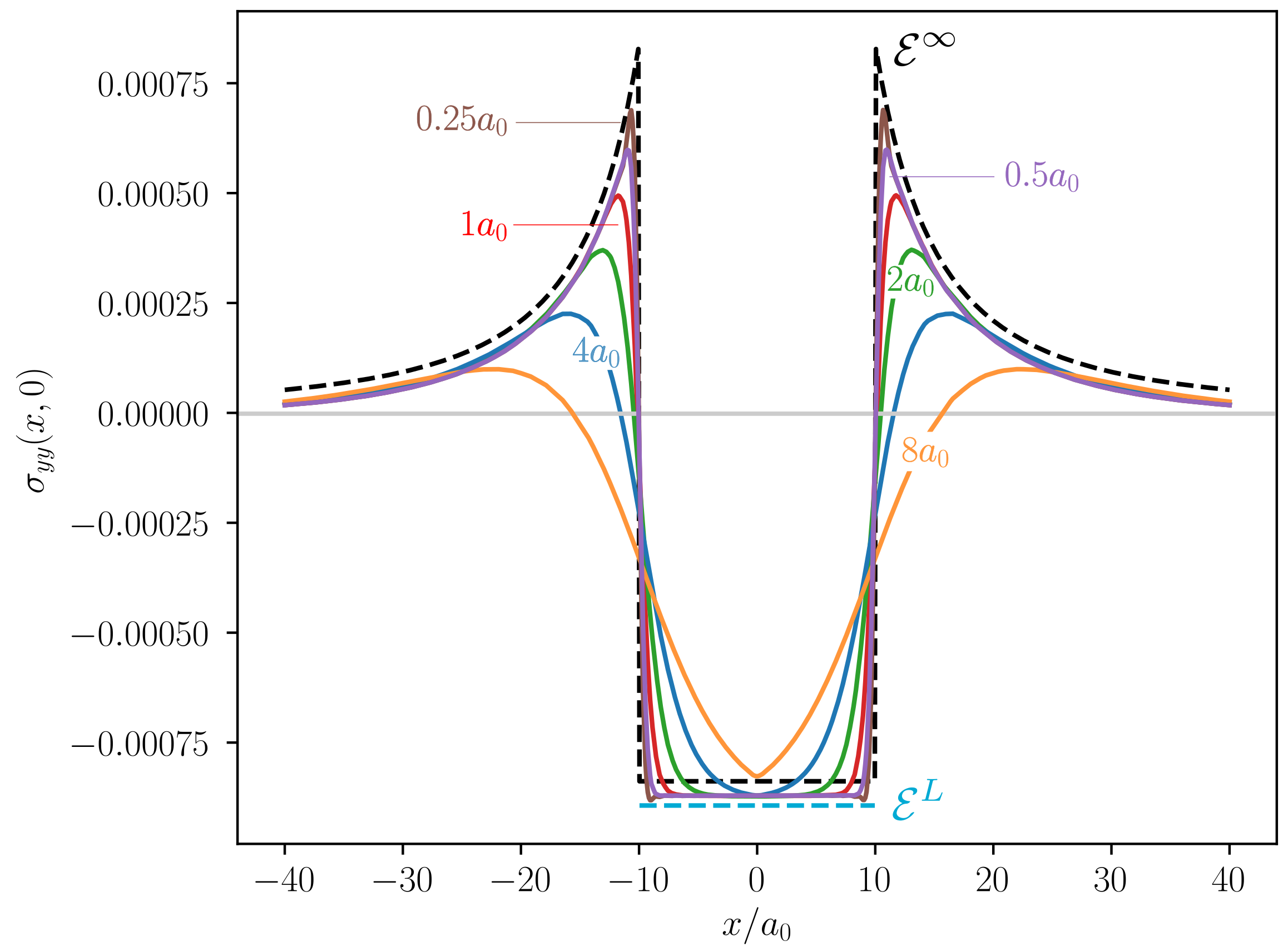}
  \caption{$\sigma_{yy}(x,0)$ for different values of $\thickness \in [0.25a_0;8a_0]$ entering eq.\ \eqref{eq:tanh} (other parameters as in Fig.~\ref{fig:figure1}). The stress field from the Eshelby solution for an infinite domain $(\mathcal{E}^{\infty})$ from eq.\ \eqref{eq:eshelby} and for a finite (circular) domain with Dirichlet boundary condition from Ref.~\cite{li2005circular} $(\mathcal{E}^L)$ with domain size $L=100a_0$.}
  \label{fig:figure2}
\end{figure}

\begin{figure}
\centering
\includegraphics[width=0.45\textwidth]{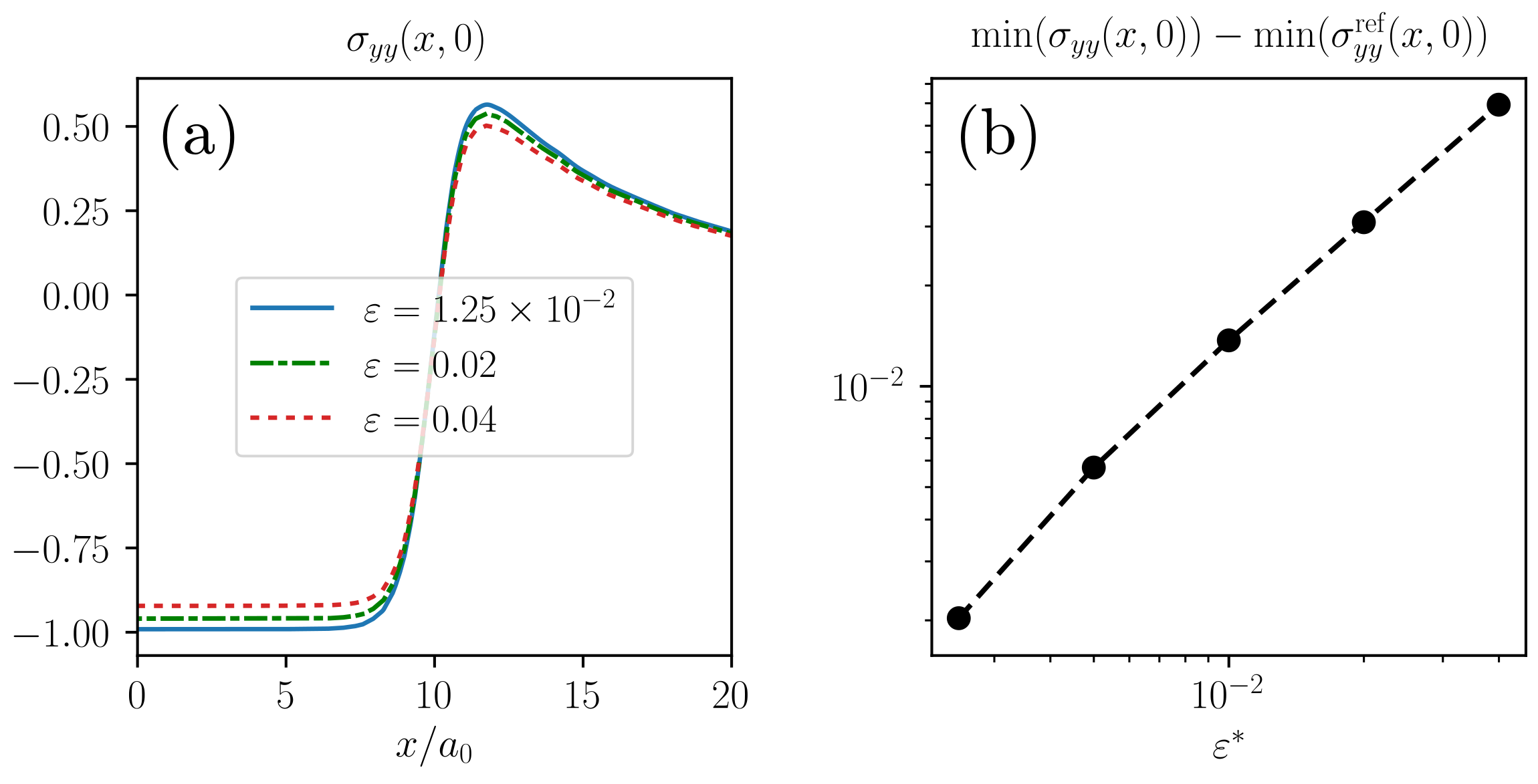}
  \caption{
Deviation from linear elasticity. (a) Normalized stress field $\sigma_{yy}(x,0)$ for different $\varepsilon^*$. (b) Deviation of the minimum of $\sigma_{yy}$ as function of $\varepsilon^*$ from the reference value ($\sigma_{yy}^{\rm ref}$) obtained with $\varepsilon^*=10^{-5}$.}
\label{fig:figure3}
\end{figure}

Deeper insights on this comparison and the role played by $\thickness$ are shown in Fig.~\ref{fig:figure2} in terms of the ${\sigma}_{yy}$ component extracted along the $x-$direction crossing the inclusion in its center. A progressively sharper stress field transition across the inclusion boundary is obtained by decreasing this parameter, approaching the continuum solution. From a quantitative point of view, minor deviations are observed for the decay far away from the inclusion and for the exact stress value in the inclusion, which may be ascribed to different contributions. First, periodic boundary conditions adopted in the simulations are not considered in the analytical solution. A good convergence to a numerical solution with increasing the domain size is obtained for the considered ratio of 0.1 among the radius of the inclusion and the side of the square simulation domain. However, boundary conditions may still affect the solution everywhere. Indeed, if considering a different analytical solution accounting for Dirichlet boundary conditions for a circular domain as in Ref.~\cite{li2005circular}, a (small) difference is obtained in the inclusion (see $\mathcal{E}^L$ values in Fig.~\ref{fig:figure2}) once setting the radius of the circular domain to $L=50a_0$. Notice, however, that this solution accounts for a different domain shape. 
Second, the (A)PFC model naturally encodes elasticity contributions beyond classical linear elasticity, namely non-linearities, strain gradient terms, and anisotropies \cite{Huter2016}. The latter should be considered generally, but for the example reported here they don't play a role as the triangular lattice has isotropic elastic constants. However, other deviations from linear elasticity are still expected. This is further illustrated in Fig.~\ref{fig:figure3}. The stress field obtained with different $\varepsilon^*$ is normalized w.r.t to the minimum value of a reference case with $\varepsilon=10^{-5}$. The deviation from the normalized curve increases with increasing eigenstrain (up to $\sim 15\%$ for maximum and minimum values of the considered stress component). Notice that due to the linear elasticity underlying eq.\ \eqref{eq:eshelby}, the corresponding normalized curves would coincide as $\varepsilon^*$ enters as a factor only. Numerical convergence to a limiting normalized curve is achieved for $\varepsilon^*\rightarrow 0$. Fig.~\ref{fig:figure3}(b) shows such a behavior for what concerns the minima of $\sigma_{yy}$. A very similar convergence behavior is obtained for the maxima of $\sigma_{yy}$.

\section{Conclusion}

In this work, we presented an (A)PFC formulation encoding an eigenstrain. This is achieved by acting on the quantity entering the free energy, which controls the equilibrium lattice parameter. The model has been benchmarked against the prototypical case of a mismatched inclusion, and it is found to match well with the solution of the Eshelby problem. Deviation from the classical analytical solutions may be ascribed to the considered simulation setup and the more detailed elasticity description conveyed by the APFC model.

The model formulation and the example of an elastic inclusion set the ground for the coarse-grained modeling of crystalline material involving mechanical deformation not caused by external mechanical stress (namely eigenstrains \cite{Kinoshita1971}). Examples of potential applications are heterostructures undergoing thermal treatment and experiencing thermal expansion \cite{Huttin2012}, heteroepitaxial systems \cite{Bergamaschini2016b}, prestretched crystalline domains \cite{Chen2020} and the already mentioned lithiumization in lithium-ion batteries \cite{Huttin2012}. 
The approach provides a possibility to consider the effect of eigenstrain in (A)PFC models. Future work will focus on exploiting the capabilities of the model in describing different lattice symmetries and three-dimensional systems as well as specific, technologically relevant applications. Various possibilities to enforce different lattice symmetries in PFC have been compared in \cite{BSWV:2020}. Our approach can be adapted to all of them. Also, the considered setting, including the approximation of the characteristic function of the inclusion in eq.\ \eqref{eq:tanh}, already provides the grounds for dynamic couplings with classical phase-field models, as e.g. considered in \cite{BalakrishnaCarter:2018} using a Cahn-Hilliard-PFC model for diffusion-induced grain boundary migration. 

\section*{Acknowledgments}
M. Salvalaglio acknowledges support from the Emmy Noether Programme of the German Research Foundation (DFG) under Grant SA4032/2-1. K. Chockalingam gratefully acknowledges financial support by the German
Research Foundation (DFG) through RTG 2218 Simulation of
Mechano-Electro-Thermal processes in Lithium-Ion-Batteries (SiMET), project
number 281041241. We gratefully acknowledge computing time grants from the Centre for Information Services and High-Performance Computing (ZIH) at TU Dresden and computing resources provided at J\"ulich
Supercomputing Center under Grant PFAMDIS.
%

\begin{thebibliography}{39}%
\makeatletter
\providecommand \@ifxundefined [1]{%
 \@ifx{#1\undefined}
}%
\providecommand \@ifnum [1]{%
 \ifnum #1\expandafter \@firstoftwo
 \else \expandafter \@secondoftwo
 \fi
}%
\providecommand \@ifx [1]{%
 \ifx #1\expandafter \@firstoftwo
 \else \expandafter \@secondoftwo
 \fi
}%
\providecommand \natexlab [1]{#1}%
\providecommand \enquote  [1]{``#1''}%
\providecommand \bibnamefont  [1]{#1}%
\providecommand \bibfnamefont [1]{#1}%
\providecommand \citenamefont [1]{#1}%
\providecommand \href@noop [0]{\@secondoftwo}%
\providecommand \href [0]{\begingroup \@sanitize@url \@href}%
\providecommand \@href[1]{\@@startlink{#1}\@@href}%
\providecommand \@@href[1]{\endgroup#1\@@endlink}%
\providecommand \@sanitize@url [0]{\catcode `\\12\catcode `\$12\catcode
  `\&12\catcode `\#12\catcode `\^12\catcode `\_12\catcode `\%12\relax}%
\providecommand \@@startlink[1]{}%
\providecommand \@@endlink[0]{}%
\providecommand \url  [0]{\begingroup\@sanitize@url \@url }%
\providecommand \@url [1]{\endgroup\@href {#1}{\urlprefix }}%
\providecommand \urlprefix  [0]{URL }%
\providecommand \Eprint [0]{\href }%
\providecommand \doibase [0]{http://dx.doi.org/}%
\providecommand \selectlanguage [0]{\@gobble}%
\providecommand \bibinfo  [0]{\@secondoftwo}%
\providecommand \bibfield  [0]{\@secondoftwo}%
\providecommand \translation [1]{[#1]}%
\providecommand \BibitemOpen [0]{}%
\providecommand \bibitemStop [0]{}%
\providecommand \bibitemNoStop [0]{.\EOS\space}%
\providecommand \EOS [0]{\spacefactor3000\relax}%
\providecommand \BibitemShut  [1]{\csname bibitem#1\endcsname}%
\let\auto@bib@innerbib\@empty
\bibitem [{\citenamefont {Li}\ and\ \citenamefont
  {Wang}(2008)}]{li2008introduction}%
  \BibitemOpen
  \bibfield  {author} {\bibinfo {author} {\bibfnamefont {S.}~\bibnamefont
  {Li}}\ and\ \bibinfo {author} {\bibfnamefont {G.}~\bibnamefont {Wang}},\
  }\href@noop {} {\emph {\bibinfo {title} {Introduction to micromechanics and
  nanomechanics}}}\ (\bibinfo  {publisher} {World Scientific Publishing
  Company},\ \bibinfo {year} {2008})\BibitemShut {NoStop}%
\bibitem [{\citenamefont {Cai}\ and\ \citenamefont
  {Nix}(2016)}]{cai2016imperfections}%
  \BibitemOpen
  \bibfield  {author} {\bibinfo {author} {\bibfnamefont {W.}~\bibnamefont
  {Cai}}\ and\ \bibinfo {author} {\bibfnamefont {W.~D.}\ \bibnamefont {Nix}},\
  }\href@noop {} {\emph {\bibinfo {title} {Imperfections in crystalline
  solids}}}\ (\bibinfo  {publisher} {Cambridge University Press},\ \bibinfo
  {year} {2016})\BibitemShut {NoStop}%
\bibitem [{\citenamefont {Bergamaschini}\ \emph {et~al.}(2016)\citenamefont
  {Bergamaschini}, \citenamefont {Salvalaglio}, \citenamefont {Backofen},
  \citenamefont {Voigt},\ and\ \citenamefont
  {Montalenti}}]{Bergamaschini2016b}%
  \BibitemOpen
  \bibfield  {author} {\bibinfo {author} {\bibfnamefont {R.}~\bibnamefont
  {Bergamaschini}}, \bibinfo {author} {\bibfnamefont {M.}~\bibnamefont
  {Salvalaglio}}, \bibinfo {author} {\bibfnamefont {R.}~\bibnamefont
  {Backofen}}, \bibinfo {author} {\bibfnamefont {A.}~\bibnamefont {Voigt}}, \
  and\ \bibinfo {author} {\bibfnamefont {F.}~\bibnamefont {Montalenti}},\
  }\href {\doibase 10.1080/23746149.2016.1181986} {\bibfield  {journal}
  {\bibinfo  {journal} {Advances in Physics: X}\ }\textbf {\bibinfo {volume}
  {1}},\ \bibinfo {pages} {331} (\bibinfo {year} {2016})}\BibitemShut {NoStop}%
\bibitem [{\citenamefont {M{\"u}ller}\ and\ \citenamefont
  {Sa{\'u}l}(2004)}]{muller2004elastic}%
  \BibitemOpen
  \bibfield  {author} {\bibinfo {author} {\bibfnamefont {P.}~\bibnamefont
  {M{\"u}ller}}\ and\ \bibinfo {author} {\bibfnamefont {A.}~\bibnamefont
  {Sa{\'u}l}},\ }\href {\doibase 10.1016/j.surfrep.2004.05.001} {\bibfield
  {journal} {\bibinfo  {journal} {Surf. Sci. Rep.}\ }\textbf {\bibinfo {volume}
  {54}},\ \bibinfo {pages} {157} (\bibinfo {year} {2004})}\BibitemShut
  {NoStop}%
\bibitem [{\citenamefont {Huttin}\ and\ \citenamefont
  {Kamlah}(2012)}]{Huttin2012}%
  \BibitemOpen
  \bibfield  {author} {\bibinfo {author} {\bibfnamefont {M.}~\bibnamefont
  {Huttin}}\ and\ \bibinfo {author} {\bibfnamefont {M.}~\bibnamefont
  {Kamlah}},\ }\href {\doibase 10.1063/1.4754705} {\bibfield  {journal}
  {\bibinfo  {journal} {Appl. Phys. Lett.}\ }\textbf {\bibinfo {volume}
  {101}},\ \bibinfo {pages} {133902} (\bibinfo {year} {2012})}\BibitemShut
  {NoStop}%
\bibitem [{\citenamefont {Elder}\ \emph {et~al.}(2002)\citenamefont {Elder},
  \citenamefont {Katakowski}, \citenamefont {Haataja},\ and\ \citenamefont
  {Grant}}]{Elder2002}%
  \BibitemOpen
  \bibfield  {author} {\bibinfo {author} {\bibfnamefont {K.~R.}\ \bibnamefont
  {Elder}}, \bibinfo {author} {\bibfnamefont {M.}~\bibnamefont {Katakowski}},
  \bibinfo {author} {\bibfnamefont {M.}~\bibnamefont {Haataja}}, \ and\
  \bibinfo {author} {\bibfnamefont {M.}~\bibnamefont {Grant}},\ }\href
  {\doibase 10.1103/PhysRevE.70.051605} {\bibfield  {journal} {\bibinfo
  {journal} {Phys. Rev. Lett.}\ }\textbf {\bibinfo {volume} {88}},\ \bibinfo
  {pages} {245701} (\bibinfo {year} {2002})}\BibitemShut {NoStop}%
\bibitem [{\citenamefont {Elder}\ and\ \citenamefont
  {Grant}(2004)}]{Elder2004}%
  \BibitemOpen
  \bibfield  {author} {\bibinfo {author} {\bibfnamefont {K.~R.}\ \bibnamefont
  {Elder}}\ and\ \bibinfo {author} {\bibfnamefont {M.}~\bibnamefont {Grant}},\
  }\href {\doibase 10.1103/PhysRevLett.88.245701} {\bibfield  {journal}
  {\bibinfo  {journal} {Phys. Rev. E}\ }\textbf {\bibinfo {volume} {70}},\
  \bibinfo {pages} {051605} (\bibinfo {year} {2004})}\BibitemShut {NoStop}%
\bibitem [{\citenamefont {Emmerich}\ \emph {et~al.}(2012)\citenamefont
  {Emmerich}, \citenamefont {L{\"{o}}wen}, \citenamefont {Wittkowski},
  \citenamefont {Gruhn}, \citenamefont {T{\'{o}}th}, \citenamefont {Tegze},\
  and\ \citenamefont {Gr{\'{a}}n{\'{a}}sy}}]{Emmerich2012}%
  \BibitemOpen
  \bibfield  {author} {\bibinfo {author} {\bibfnamefont {H.}~\bibnamefont
  {Emmerich}}, \bibinfo {author} {\bibfnamefont {H.}~\bibnamefont
  {L{\"{o}}wen}}, \bibinfo {author} {\bibfnamefont {R.}~\bibnamefont
  {Wittkowski}}, \bibinfo {author} {\bibfnamefont {T.}~\bibnamefont {Gruhn}},
  \bibinfo {author} {\bibfnamefont {G.~I.}\ \bibnamefont {T{\'{o}}th}},
  \bibinfo {author} {\bibfnamefont {G.}~\bibnamefont {Tegze}}, \ and\ \bibinfo
  {author} {\bibfnamefont {L.}~\bibnamefont {Gr{\'{a}}n{\'{a}}sy}},\ }\href
  {\doibase 10.1080/00018732.2012.737555} {\bibfield  {journal} {\bibinfo
  {journal} {Advances in Physics}\ }\textbf {\bibinfo {volume} {61}},\ \bibinfo
  {pages} {665} (\bibinfo {year} {2012})}\BibitemShut {NoStop}%
\bibitem [{\citenamefont {Goldenfeld}\ \emph {et~al.}(2005)\citenamefont
  {Goldenfeld}, \citenamefont {Athreya},\ and\ \citenamefont
  {Dantzig}}]{Goldenfeld2005}%
  \BibitemOpen
  \bibfield  {author} {\bibinfo {author} {\bibfnamefont {N.}~\bibnamefont
  {Goldenfeld}}, \bibinfo {author} {\bibfnamefont {B.~P.}\ \bibnamefont
  {Athreya}}, \ and\ \bibinfo {author} {\bibfnamefont {J.~A.}\ \bibnamefont
  {Dantzig}},\ }\href {\doibase 10.1103/PhysRevE.72.020601} {\bibfield
  {journal} {\bibinfo  {journal} {Phys. Rev. E}\ }\textbf {\bibinfo {volume}
  {72}},\ \bibinfo {pages} {020601} (\bibinfo {year} {2005})}\BibitemShut
  {NoStop}%
\bibitem [{\citenamefont {Athreya}\ \emph {et~al.}(2006)\citenamefont
  {Athreya}, \citenamefont {Goldenfeld},\ and\ \citenamefont
  {Dantzig}}]{Athreya2006}%
  \BibitemOpen
  \bibfield  {author} {\bibinfo {author} {\bibfnamefont {B.~P.}\ \bibnamefont
  {Athreya}}, \bibinfo {author} {\bibfnamefont {N.}~\bibnamefont {Goldenfeld}},
  \ and\ \bibinfo {author} {\bibfnamefont {J.~A.}\ \bibnamefont {Dantzig}},\
  }\href {\doibase 10.1103/PhysRevE.74.011601} {\bibfield  {journal} {\bibinfo
  {journal} {Phys. Rev. E}\ }\textbf {\bibinfo {volume} {74}},\ \bibinfo
  {pages} {011601} (\bibinfo {year} {2006})}\BibitemShut {NoStop}%
\bibitem [{\citenamefont {Yeon}\ \emph {et~al.}(2010)\citenamefont {Yeon},
  \citenamefont {Huang}, \citenamefont {Elder},\ and\ \citenamefont
  {Thornton}}]{YHET:2010}%
  \BibitemOpen
  \bibfield  {author} {\bibinfo {author} {\bibfnamefont {D.-H.}\ \bibnamefont
  {Yeon}}, \bibinfo {author} {\bibfnamefont {Z.-F.}\ \bibnamefont {Huang}},
  \bibinfo {author} {\bibfnamefont {K.~R.}\ \bibnamefont {Elder}}, \ and\
  \bibinfo {author} {\bibfnamefont {K.}~\bibnamefont {Thornton}},\ }\href
  {\doibase 10.1080/14786430903164572} {\bibfield  {journal} {\bibinfo
  {journal} {Philosophical Magazine}\ }\textbf {\bibinfo {volume} {90}},\
  \bibinfo {pages} {1} (\bibinfo {year} {2010})}\BibitemShut {NoStop}%
\bibitem [{\citenamefont {Salvalaglio}\ and\ \citenamefont
  {Elder}(2022)}]{SE:2022}%
  \BibitemOpen
  \bibfield  {author} {\bibinfo {author} {\bibfnamefont {M.}~\bibnamefont
  {Salvalaglio}}\ and\ \bibinfo {author} {\bibfnamefont {R.~K.}\ \bibnamefont
  {Elder}},\ }\href@noop {} {\bibfield  {journal} {\bibinfo  {journal} {Review
  article: \href{https://arxiv.org/abs/2202.04196}{\textit{Coarse-grained Modeling of Crystals by the Amplitude
  Expansion of the Phase-Field Crystal Model: an Overview}}, under revision}\ }
  (\bibinfo {year} {2022})}\BibitemShut {NoStop}%
\bibitem [{\citenamefont {Elder}\ \emph {et~al.}(2010)\citenamefont {Elder},
  \citenamefont {Huang},\ and\ \citenamefont {Provatas}}]{ElderPRE2010}%
  \BibitemOpen
  \bibfield  {author} {\bibinfo {author} {\bibfnamefont {K.~R.}\ \bibnamefont
  {Elder}}, \bibinfo {author} {\bibfnamefont {Z.-F.}\ \bibnamefont {Huang}}, \
  and\ \bibinfo {author} {\bibfnamefont {N.}~\bibnamefont {Provatas}},\ }\href
  {\doibase 10.1103/PhysRevE.81.011602} {\bibfield  {journal} {\bibinfo
  {journal} {Phys. Rev. E}\ }\textbf {\bibinfo {volume} {81}},\ \bibinfo
  {pages} {011602} (\bibinfo {year} {2010})}\BibitemShut {NoStop}%
\bibitem [{\citenamefont {Spatschek}\ and\ \citenamefont
  {Karma}(2010)}]{SpatschekKarma:2010}%
  \BibitemOpen
  \bibfield  {author} {\bibinfo {author} {\bibfnamefont {R.}~\bibnamefont
  {Spatschek}}\ and\ \bibinfo {author} {\bibfnamefont {A.}~\bibnamefont
  {Karma}},\ }\href {\doibase 10.1103/PhysRevB.81.214201} {\bibfield  {journal}
  {\bibinfo  {journal} {Phys. Rev. B}\ }\textbf {\bibinfo {volume} {81}},\
  \bibinfo {pages} {214201} (\bibinfo {year} {2010})}\BibitemShut {NoStop}%
\bibitem [{\citenamefont {Heinonen}\ \emph {et~al.}(2014)\citenamefont
  {Heinonen}, \citenamefont {Achim}, \citenamefont {Elder}, \citenamefont
  {Buyukdagli},\ and\ \citenamefont {Ala-Nissila}}]{Heinonen2014}%
  \BibitemOpen
  \bibfield  {author} {\bibinfo {author} {\bibfnamefont {V.}~\bibnamefont
  {Heinonen}}, \bibinfo {author} {\bibfnamefont {C.~V.}\ \bibnamefont {Achim}},
  \bibinfo {author} {\bibfnamefont {K.~R.}\ \bibnamefont {Elder}}, \bibinfo
  {author} {\bibfnamefont {S.}~\bibnamefont {Buyukdagli}}, \ and\ \bibinfo
  {author} {\bibfnamefont {T.}~\bibnamefont {Ala-Nissila}},\ }\href {\doibase
  10.1103/PhysRevE.89.032411} {\bibfield  {journal} {\bibinfo  {journal} {Phys.
  Rev. E}\ }\textbf {\bibinfo {volume} {89}},\ \bibinfo {pages} {032411}
  (\bibinfo {year} {2014})}\BibitemShut {NoStop}%
\bibitem [{\citenamefont {H{\"{u}}ter}\ \emph {et~al.}(2016)\citenamefont
  {H{\"{u}}ter}, \citenamefont {Fri{\'{a}}k}, \citenamefont {Weikamp},
  \citenamefont {Neugebauer}, \citenamefont {Goldenfeld}, \citenamefont
  {Svendsen},\ and\ \citenamefont {Spatschek}}]{Huter2016}%
  \BibitemOpen
  \bibfield  {author} {\bibinfo {author} {\bibfnamefont {C.}~\bibnamefont
  {H{\"{u}}ter}}, \bibinfo {author} {\bibfnamefont {M.}~\bibnamefont
  {Fri{\'{a}}k}}, \bibinfo {author} {\bibfnamefont {M.}~\bibnamefont
  {Weikamp}}, \bibinfo {author} {\bibfnamefont {J.}~\bibnamefont {Neugebauer}},
  \bibinfo {author} {\bibfnamefont {N.}~\bibnamefont {Goldenfeld}}, \bibinfo
  {author} {\bibfnamefont {B.}~\bibnamefont {Svendsen}}, \ and\ \bibinfo
  {author} {\bibfnamefont {R.}~\bibnamefont {Spatschek}},\ }\href {\doibase
  10.1103/PhysRevB.93.214105} {\bibfield  {journal} {\bibinfo  {journal} {Phys.
  Rev. B}\ }\textbf {\bibinfo {volume} {93}},\ \bibinfo {pages} {214105}
  (\bibinfo {year} {2016})}\BibitemShut {NoStop}%
\bibitem [{\citenamefont {Salvalaglio}\ \emph {et~al.}(2019)\citenamefont
  {Salvalaglio}, \citenamefont {Voigt},\ and\ \citenamefont
  {Elder}}]{SalvalaglioNPJ2019}%
  \BibitemOpen
  \bibfield  {author} {\bibinfo {author} {\bibfnamefont {M.}~\bibnamefont
  {Salvalaglio}}, \bibinfo {author} {\bibfnamefont {A.}~\bibnamefont {Voigt}},
  \ and\ \bibinfo {author} {\bibfnamefont {K.~R.}\ \bibnamefont {Elder}},\
  }\href {\doibase 10.1038/s41524-019-0185-0} {\bibfield  {journal} {\bibinfo
  {journal} {npj Computational Materials}\ }\textbf {\bibinfo {volume} {5}},\
  \bibinfo {pages} {48} (\bibinfo {year} {2019})}\BibitemShut {NoStop}%
\bibitem [{\citenamefont {Salvalaglio}\ \emph {et~al.}(2020)\citenamefont
  {Salvalaglio}, \citenamefont {Angheluta}, \citenamefont {Huang},
  \citenamefont {Voigt}, \citenamefont {Elder},\ and\ \citenamefont
  {Vi{\~n}als}}]{SAHVEV:2020}%
  \BibitemOpen
  \bibfield  {author} {\bibinfo {author} {\bibfnamefont {M.}~\bibnamefont
  {Salvalaglio}}, \bibinfo {author} {\bibfnamefont {L.}~\bibnamefont
  {Angheluta}}, \bibinfo {author} {\bibfnamefont {Z.-F.}\ \bibnamefont
  {Huang}}, \bibinfo {author} {\bibfnamefont {A.}~\bibnamefont {Voigt}},
  \bibinfo {author} {\bibfnamefont {K.~R.}\ \bibnamefont {Elder}}, \ and\
  \bibinfo {author} {\bibfnamefont {J.}~\bibnamefont {Vi{\~n}als}},\ }\href
  {\doibase https://doi.org/10.1016/j.jmps.2019.103856} {\bibfield  {journal}
  {\bibinfo  {journal} {J. Mech. Phys. Solids}\ }\textbf {\bibinfo {volume}
  {137}},\ \bibinfo {pages} {103856} (\bibinfo {year} {2020})}\BibitemShut
  {NoStop}%
\bibitem [{\citenamefont {Salvalaglio}\ \emph {et~al.}(2021)\citenamefont
  {Salvalaglio}, \citenamefont {Voigt}, \citenamefont {Huang},\ and\
  \citenamefont {Elder}}]{SalvalaglioPRL2021}%
  \BibitemOpen
  \bibfield  {author} {\bibinfo {author} {\bibfnamefont {M.}~\bibnamefont
  {Salvalaglio}}, \bibinfo {author} {\bibfnamefont {A.}~\bibnamefont {Voigt}},
  \bibinfo {author} {\bibfnamefont {Z.-F.}\ \bibnamefont {Huang}}, \ and\
  \bibinfo {author} {\bibfnamefont {K.~R.}\ \bibnamefont {Elder}},\ }\href
  {\doibase 10.1103/PhysRevLett.126.185502} {\bibfield  {journal} {\bibinfo
  {journal} {Phys. Rev. Lett.}\ }\textbf {\bibinfo {volume} {126}},\ \bibinfo
  {pages} {185502} (\bibinfo {year} {2021})}\BibitemShut {NoStop}%
\bibitem [{\citenamefont {Kinoshita}\ and\ \citenamefont
  {Mura}(1971)}]{Kinoshita1971}%
  \BibitemOpen
  \bibfield  {author} {\bibinfo {author} {\bibfnamefont {N.}~\bibnamefont
  {Kinoshita}}\ and\ \bibinfo {author} {\bibfnamefont {T.}~\bibnamefont
  {Mura}},\ }\href {\doibase 10.1002/pssa.2210050332} {\bibfield  {journal}
  {\bibinfo  {journal} {physica status solidi (a)}\ }\textbf {\bibinfo {volume}
  {5}},\ \bibinfo {pages} {759} (\bibinfo {year} {1971})}\BibitemShut {NoStop}%
\bibitem [{\citenamefont {Eshelby}(1957)}]{eshelby1957determination}%
  \BibitemOpen
  \bibfield  {author} {\bibinfo {author} {\bibfnamefont {J.~D.}\ \bibnamefont
  {Eshelby}},\ }\href {\doibase 10.1098/rspa.1957.0133} {\bibfield  {journal}
  {\bibinfo  {journal} {Proc. Roy. Soc. London. Ser. A. Math. Phys. Sci.}\
  }\textbf {\bibinfo {volume} {241}},\ \bibinfo {pages} {376} (\bibinfo {year}
  {1957})}\BibitemShut {NoStop}%
\bibitem [{\citenamefont {Eshelby}(1959)}]{eshelby1959elastic}%
  \BibitemOpen
  \bibfield  {author} {\bibinfo {author} {\bibfnamefont {J.~D.}\ \bibnamefont
  {Eshelby}},\ }\href {\doibase 10.1098/rspa.1959.0173} {\bibfield  {journal}
  {\bibinfo  {journal} {Proc. Roy. Soc. London. Ser. A. Math. Phys. Sci.}\
  }\textbf {\bibinfo {volume} {252}},\ \bibinfo {pages} {561} (\bibinfo {year}
  {1959})}\BibitemShut {NoStop}%
\bibitem [{\citenamefont {Mura}(2013)}]{mura2013micromechanics}%
  \BibitemOpen
  \bibfield  {author} {\bibinfo {author} {\bibfnamefont {T.}~\bibnamefont
  {Mura}},\ }\href@noop {} {\emph {\bibinfo {title} {Micromechanics of defects
  in solids}}}\ (\bibinfo  {publisher} {Springer Science \& Business Media},\
  \bibinfo {year} {2013})\BibitemShut {NoStop}%
\bibitem [{\citenamefont {Elder}\ \emph {et~al.}(2007)\citenamefont {Elder},
  \citenamefont {Provatas}, \citenamefont {Berry}, \citenamefont {Stefanovic},\
  and\ \citenamefont {Grant}}]{elder2007}%
  \BibitemOpen
  \bibfield  {author} {\bibinfo {author} {\bibfnamefont {K.~R.}\ \bibnamefont
  {Elder}}, \bibinfo {author} {\bibfnamefont {N.}~\bibnamefont {Provatas}},
  \bibinfo {author} {\bibfnamefont {J.}~\bibnamefont {Berry}}, \bibinfo
  {author} {\bibfnamefont {P.}~\bibnamefont {Stefanovic}}, \ and\ \bibinfo
  {author} {\bibfnamefont {M.}~\bibnamefont {Grant}},\ }\href{\doibase 10.1103/PhysRevB.75.064107}
  {\bibfield  {journal} {\bibinfo  {journal} {Phys. Rev. B}\ }\textbf {\bibinfo
  {volume} {75}},\ \bibinfo {pages} {064107} (\bibinfo {year}
  {2007})}\BibitemShut {NoStop}%
\bibitem [{\citenamefont {Salvalaglio}\ \emph {et~al.}(2017)\citenamefont
  {Salvalaglio}, \citenamefont {Backofen}, \citenamefont {Voigt},\ and\
  \citenamefont {Elder}}]{SBVE:2017}%
  \BibitemOpen
  \bibfield  {author} {\bibinfo {author} {\bibfnamefont {M.}~\bibnamefont
  {Salvalaglio}}, \bibinfo {author} {\bibfnamefont {R.}~\bibnamefont
  {Backofen}}, \bibinfo {author} {\bibfnamefont {A.}~\bibnamefont {Voigt}}, \
  and\ \bibinfo {author} {\bibfnamefont {K.}~\bibnamefont {Elder}},\ }\href
  {\doibase 10.1103/PhysRevE.96.023301} {\bibfield  {journal} {\bibinfo
  {journal} {Phys. Rev. E}\ }\textbf {\bibinfo {volume} {96}},\ \bibinfo
  {pages} {023301} (\bibinfo {year} {2017})}\BibitemShut {NoStop}%
\bibitem [{\citenamefont {Skaugen}\ \emph
  {et~al.}(2018{\natexlab{a}})\citenamefont {Skaugen}, \citenamefont
  {Angheluta},\ and\ \citenamefont {Vi{\~{n}}als}}]{Skaugen2018}%
  \BibitemOpen
  \bibfield  {author} {\bibinfo {author} {\bibfnamefont {A.}~\bibnamefont
  {Skaugen}}, \bibinfo {author} {\bibfnamefont {L.}~\bibnamefont {Angheluta}},
  \ and\ \bibinfo {author} {\bibfnamefont {J.}~\bibnamefont {Vi{\~{n}}als}},\
  }\href {http://arxiv.org/abs/1710.05320
  https://link.aps.org/doi/10.1103/PhysRevB.97.054113} {\bibfield  {journal}
  {\bibinfo  {journal} {Phys. Rev. B}\ }\textbf {\bibinfo {volume} {97}},\
  \bibinfo {pages} {054113} (\bibinfo {year} {2018}{\natexlab{a}})}\BibitemShut
  {NoStop}%
\bibitem [{\citenamefont {Skaugen}\ \emph
  {et~al.}(2018{\natexlab{b}})\citenamefont {Skaugen}, \citenamefont
  {Angheluta},\ and\ \citenamefont {Vi{\~{n}}als}}]{Skaugen2018b}%
  \BibitemOpen
  \bibfield  {author} {\bibinfo {author} {\bibfnamefont {A.}~\bibnamefont
  {Skaugen}}, \bibinfo {author} {\bibfnamefont {L.}~\bibnamefont {Angheluta}},
  \ and\ \bibinfo {author} {\bibfnamefont {J.}~\bibnamefont {Vi{\~{n}}als}},\
  }\href {\doibase 10.1103/PhysRevLett.121.255501} {\bibfield  {journal}
  {\bibinfo  {journal} {Phys Rev. Lett.}\ }\textbf {\bibinfo {volume} {121}},\
  \bibinfo {pages} {255501} (\bibinfo {year} {2018}{\natexlab{b}})}\BibitemShut
  {NoStop}%
\bibitem [{\citenamefont {Skogvoll}\ \emph {et~al.}(2021)\citenamefont
  {Skogvoll}, \citenamefont {Skaugen}, \citenamefont {Angheluta},\ and\
  \citenamefont {Vi\~nals}}]{SSAV:2021}%
  \BibitemOpen
  \bibfield  {author} {\bibinfo {author} {\bibfnamefont {V.}~\bibnamefont
  {Skogvoll}}, \bibinfo {author} {\bibfnamefont {A.}~\bibnamefont {Skaugen}},
  \bibinfo {author} {\bibfnamefont {L.}~\bibnamefont {Angheluta}}, \ and\
  \bibinfo {author} {\bibfnamefont {J.}~\bibnamefont {Vi\~nals}},\ }\href
  {\doibase 10.1103/PhysRevB.103.014107} {\bibfield  {journal} {\bibinfo
  {journal} {Phys. Rev. B}\ }\textbf {\bibinfo {volume} {103}},\ \bibinfo
  {pages} {014107} (\bibinfo {year} {2021})}\BibitemShut {NoStop}%
\bibitem [{\citenamefont {Ju}\ and\ \citenamefont {Sun}(1999)}]{ju1999novel}%
  \BibitemOpen
  \bibfield  {author} {\bibinfo {author} {\bibfnamefont {J.}~\bibnamefont
  {Ju}}\ and\ \bibinfo {author} {\bibfnamefont {L.}~\bibnamefont {Sun}},\
  }\href {\doibase 10.1115/1.2791090} {\bibfield  {journal} {\bibinfo
  {journal} {J. Appl. Mech.}\ }\textbf {\bibinfo {volume} {66}},\ \bibinfo
  {pages} {570} (\bibinfo {year} {1999})}\BibitemShut {NoStop}%
\bibitem [{\citenamefont {Li}\ \emph {et~al.}(2005)\citenamefont {Li},
  \citenamefont {Sauer},\ and\ \citenamefont {Wang}}]{li2005circular}%
  \BibitemOpen
  \bibfield  {author} {\bibinfo {author} {\bibfnamefont {S.}~\bibnamefont
  {Li}}, \bibinfo {author} {\bibfnamefont {R.}~\bibnamefont {Sauer}}, \ and\
  \bibinfo {author} {\bibfnamefont {G.}~\bibnamefont {Wang}},\ }\href {\doibase
  10.1007/s00707-005-0234-2} {\bibfield  {journal} {\bibinfo  {journal} {Acta
  mechanica}\ }\textbf {\bibinfo {volume} {179}},\ \bibinfo {pages} {67}
  (\bibinfo {year} {2005})}\BibitemShut {NoStop}%
\bibitem [{\citenamefont {Wang}\ \emph {et~al.}(2005)\citenamefont {Wang},
  \citenamefont {Li},\ and\ \citenamefont {Sauer}}]{wang2005circular}%
  \BibitemOpen
  \bibfield  {author} {\bibinfo {author} {\bibfnamefont {G.}~\bibnamefont
  {Wang}}, \bibinfo {author} {\bibfnamefont {S.}~\bibnamefont {Li}}, \ and\
  \bibinfo {author} {\bibfnamefont {R.}~\bibnamefont {Sauer}},\ }\href
  {\doibase 10.1007/s00707-005-0236-0} {\bibfield  {journal} {\bibinfo
  {journal} {Acta mechanica}\ }\textbf {\bibinfo {volume} {179}},\ \bibinfo
  {pages} {91} (\bibinfo {year} {2005})}\BibitemShut {NoStop}%
\bibitem [{\citenamefont {Fischer}\ \emph {et~al.}(2018)\citenamefont
  {Fischer}, \citenamefont {Zickler},\ and\ \citenamefont
  {Svoboda}}]{fischer2018elastic}%
  \BibitemOpen
  \bibfield  {author} {\bibinfo {author} {\bibfnamefont {F.-D.}\ \bibnamefont
  {Fischer}}, \bibinfo {author} {\bibfnamefont {G.}~\bibnamefont {Zickler}}, \
  and\ \bibinfo {author} {\bibfnamefont {J.}~\bibnamefont {Svoboda}},\ }\href
  {\doibase 10.1007/s00419-017-1318-x} {\bibfield  {journal} {\bibinfo
  {journal} {Arch. Appl. Mech.}\ }\textbf {\bibinfo {volume} {88}},\ \bibinfo
  {pages} {453} (\bibinfo {year} {2018})}\BibitemShut {NoStop}%
\bibitem [{Note1()}]{Note1}%
  \BibitemOpen
  \bibinfo {note} {With (A)PFC we refer to both PFC and APFC.}\BibitemShut
  {Stop}%
\bibitem [{\citenamefont {Vey}\ and\ \citenamefont {Voigt}(2007)}]{Vey2007}%
  \BibitemOpen
  \bibfield  {author} {\bibinfo {author} {\bibfnamefont {S.}~\bibnamefont
  {Vey}}\ and\ \bibinfo {author} {\bibfnamefont {A.}~\bibnamefont {Voigt}},\
  }\href {\doibase 10.1007/s00791-006-0048-3} {\bibfield  {journal} {\bibinfo
  {journal} {Comput. Visual. Sci.}\ }\textbf {\bibinfo {volume} {10}},\
  \bibinfo {pages} {57} (\bibinfo {year} {2007})}\BibitemShut {NoStop}%
\bibitem [{\citenamefont {Witkowski}\ \emph {et~al.}(2015)\citenamefont
  {Witkowski}, \citenamefont {Ling}, \citenamefont {Praetorius},\ and\
  \citenamefont {Voigt}}]{WitkowskiACM2015}%
  \BibitemOpen
  \bibfield  {author} {\bibinfo {author} {\bibfnamefont {T.}~\bibnamefont
  {Witkowski}}, \bibinfo {author} {\bibfnamefont {S.}~\bibnamefont {Ling}},
  \bibinfo {author} {\bibfnamefont {S.}~\bibnamefont {Praetorius}}, \ and\
  \bibinfo {author} {\bibfnamefont {A.}~\bibnamefont {Voigt}},\ }\href
  {\doibase 10.1007/s10444-015-9405-4} {\bibfield  {journal} {\bibinfo
  {journal} {Adv. Comput. Math.}\ }\textbf {\bibinfo {volume} {41}},\ \bibinfo
  {pages} {1145} (\bibinfo {year} {2015})}\BibitemShut {NoStop}%
\bibitem [{\citenamefont {Praetorius}\ \emph {et~al.}(2019)\citenamefont
  {Praetorius}, \citenamefont {Salvalaglio},\ and\ \citenamefont
  {Voigt}}]{Praetorius2019}%
  \BibitemOpen
  \bibfield  {author} {\bibinfo {author} {\bibfnamefont {S.}~\bibnamefont
  {Praetorius}}, \bibinfo {author} {\bibfnamefont {M.}~\bibnamefont
  {Salvalaglio}}, \ and\ \bibinfo {author} {\bibfnamefont {A.}~\bibnamefont
  {Voigt}},\ }\href {\doibase 10.1088/1361-651X/ab1508} {\bibfield  {journal}
  {\bibinfo  {journal} {Model. Sim. Materials Sci. Eng.}\ }\textbf {\bibinfo
  {volume} {27}},\ \bibinfo {pages} {044004} (\bibinfo {year}
  {2019})}\BibitemShut {NoStop}%
\bibitem [{\citenamefont {Chen}\ \emph {et~al.}(2020)\citenamefont {Chen},
  \citenamefont {Chen}, \citenamefont {Zhang}, \citenamefont {Li},\ and\
  \citenamefont {Li}}]{Chen2020}%
  \BibitemOpen
  \bibfield  {author} {\bibinfo {author} {\bibfnamefont {S.}~\bibnamefont
  {Chen}}, \bibinfo {author} {\bibfnamefont {J.}~\bibnamefont {Chen}}, \bibinfo
  {author} {\bibfnamefont {X.}~\bibnamefont {Zhang}}, \bibinfo {author}
  {\bibfnamefont {Z.-Y.}\ \bibnamefont {Li}}, \ and\ \bibinfo {author}
  {\bibfnamefont {J.}~\bibnamefont {Li}},\ }\href {\doibase
  10.1038/s41377-020-0309-9} {\bibfield  {journal} {\bibinfo  {journal} {Light:
  Science \& Applications}\ }\textbf {\bibinfo {volume} {9}},\ \bibinfo {pages}
  {75} (\bibinfo {year} {2020})}\BibitemShut {NoStop}%
\bibitem [{\citenamefont {Backofen}\ \emph {et~al.}(2020)\citenamefont
  {Backofen}, \citenamefont {Sahlmann}, \citenamefont {Willmann},\ and\
  \citenamefont {Voigt}}]{BSWV:2020}%
  \BibitemOpen
  \bibfield  {author} {\bibinfo {author} {\bibfnamefont {R.}~\bibnamefont
  {Backofen}}, \bibinfo {author} {\bibfnamefont {L.}~\bibnamefont {Sahlmann}},
  \bibinfo {author} {\bibfnamefont {A.}~\bibnamefont {Willmann}}, \ and\
  \bibinfo {author} {\bibfnamefont {A.}~\bibnamefont {Voigt}},\ }\href{\doibase 10.1002/pamm.202000192}{\bibfield  {journal} {\bibinfo  {journal} {Proc. Appl. Math. Mech.}\
  }\textbf {\bibinfo {volume} {20}},\ \bibinfo {pages} {e2020000192} (\bibinfo
  {year} {2020})}\BibitemShut {NoStop}%
\bibitem [{\citenamefont {Balakrishna}\ and\ \citenamefont
  {Carter}(2018)}]{BalakrishnaCarter:2018}%
  \BibitemOpen
  \bibfield  {author} {\bibinfo {author} {\bibfnamefont {A.~R.}\ \bibnamefont
  {Balakrishna}}\ and\ \bibinfo {author} {\bibfnamefont {W.~C.}\ \bibnamefont
  {Carter}},\ }\href {\doibase 10.1103/PhysRevE.97.043304} {\bibfield
  {journal} {\bibinfo  {journal} {Phys. Rev. E}\ }\textbf {\bibinfo {volume}
  {97}},\ \bibinfo {pages} {043304} (\bibinfo {year} {2018})}\BibitemShut
  {NoStop}%
\end{thebibliography}
\end{document}